\documentclass[10pt]{article}
\usepackage{graphicx}% Include figure files

\usepackage{epsfig}
\usepackage{latexsym}
\usepackage{amsmath}

\begin{document}

\title{Subfield Effects on the Core of Coauthors 
}

\author{Hassan Bougrine\footnote{Corresponding address:
Biblioth\`eque des Sciences et Techniques (BST), B6a, all\' ee de la Chimie 3,  Universit\'e de Li\`ege, B-4000 Li\`ege, Belgium Tel.: +32 4 366 9845-    $e$-$mail$ $address$:h.bougrine@ulg.ac.be}    \\  
Biblioth\`eque des Sciences et Techniques (BST) \\B6a, all\' ee de la Chimie 3\\ Universit\'e de Li\`ege\\B-4000 Li\`ege, Belgium  }

 \date{\today}
\maketitle
 \vskip 0.5truecm

\begin{abstract}
It is examined whether the number ($J$) of (joint) publications of a  "main scientist"  with her/his coauthors ranked according to rank ($r$) importance, i.e.  $ J \propto 1/r $, as found by Ausloos \cite{Sofia3a} still holds for subfields, i.e. when the "main scientist" has worked on different, sometimes overlapping, subfields. Two cases are studied. It is shown that the law holds for large subfields. As shown, in an Appendix, is also useful to combine small topics into large ones for better statistics. It is observed that the sub-cores  are much smaller than the overall coauthor core measure. Nevertheless, the smallness of the core and sub-cores may imply further considerations for the evaluation  of team research purposes and activities.

 \end{abstract}

keywords : ranking; power laws; co-authorship; research topics; coauthor core
 
\maketitle
\section{Introduction  }\label{sec:intro}

     \begin{figure}
\centering
  \includegraphics[height=18.8cm,width=16.8cm]{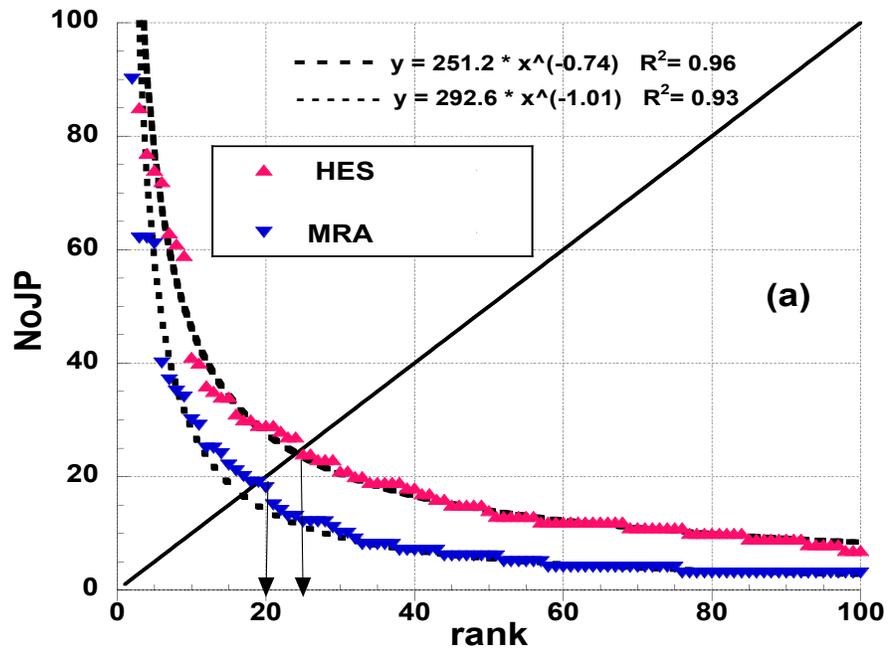}
% \newpage \includegraphics[height=18.8cm,width=9.8cm]{colPlot99bwHESMRA1b.pdf}
\caption{  Number of joint publications (NoJP) for HES and for MRA, with coauthors ranked by decreasing importance:  (a)  in the vicinity of the  so called Ausloos coauthor core measure  \cite{Sofia3a}, shown by the diagonal; values for MRA and HES are indicated by arrows;  (b)  log-log scale display  of  (a); best fits are given  for the central plot region in (a) and  (b), and also  for the  overall range in (b)}
\label{fig:MRAHESa}
\end{figure}
     \begin{figure}
\centering
  \includegraphics[height=20.8cm,width=16.8cm]{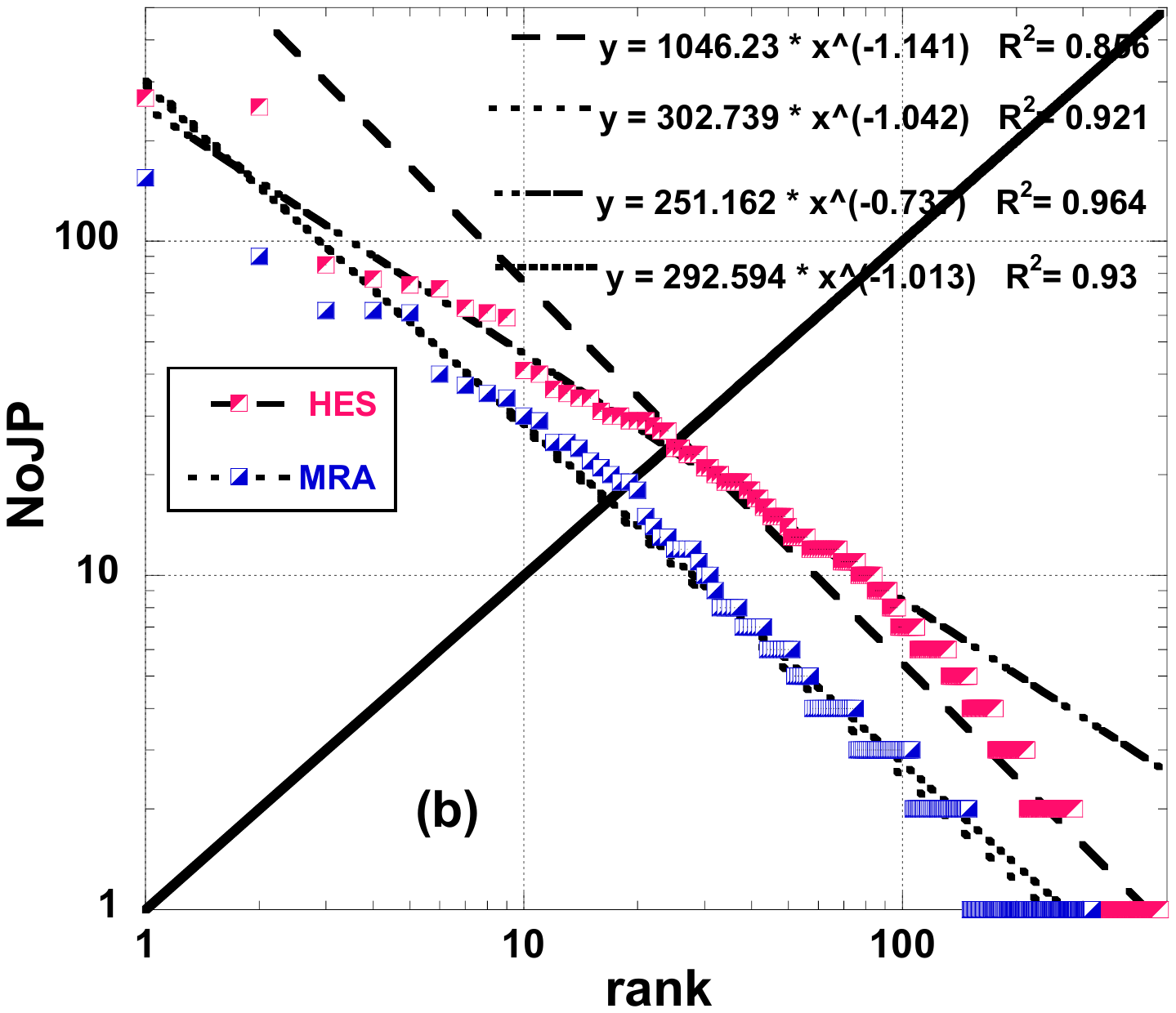}
\caption{  Number of joint publications (NoJP) for HES and for MRA, with coauthors ranked by decreasing importance:  (a)  in the vicinity of the  so called Ausloos coauthor core measure  \cite{Sofia3a}, shown by the diagonal; values for MRA and HES are indicated by arrows;  (b)  log-log scale display  of  (a); best fits are given  for the central plot region in (a) and  (b), and also  for the  overall range in (b)}
\label{fig:MRAHESb}
\end{figure}
   
Ausloos \cite{Sofia3a} has found a  simple power law relating the number of coauthors of a scientist with their rank, measured through the number of coauthored papers.  The number ($J$) of (joint) publications with coauthors ranked according to rank ($r$) importance, indicates that  $ J \propto 1/r^{\alpha}$, with $\alpha\simeq 1$. For example, comparing Ausloos (MRA) and another major scientist in statistical physics H.E. Stanley (HES) list of coauthors (more than 310 and 480 respectively) and the joint publications  (more than 560 and 870 respectively) of MRA and HES with such coauthors,  see Table 1 for a summary, one obtains a remarkable hyperbolic fit,  Figs. \ref{fig:MRAHESa}-\ref{fig:MRAHESb}, - at least in the central region. This hyperbolic law seems to be more precise when the scientist has many publications and many coauthors.   Compare the $R^2$ values for MRA and HES on Figs. 1-2. Deviations are seen mainly in the extreme regions. Moreover, the power law exponent is not exactly +1. It depends on the examined data range, see Fig. \ref{fig:MRAHESb}, - as  usual and as is well known \cite{SIAM51.09.661powerlaws} .

However, the law is interesting for two main reasons. First, instead of focussing on (the number of)  citations of papers, like for the Hirsch index  \cite{hindex,hirsch10} $h$, Ausloos focussed on  (the number of) coauthors. It has to be emphasized that this is quite different from  several variants of the $h$-index which attempted to take into account some role of coauthors for obtaining some measure of some author scientific  impact, in the literature. Next, the  approach  of Ausloos leads to some insight into team functioning.  Thus, it allows to define the {\it core  of coauthors} of a scientist, through ($m_a \equiv r =J$)  in contrast to the {\it the core  of papers} of an author, i.e. $h$. Technically, one could thus measure the relevant strength of a research group centered on some leader. The invisible college  \cite{HK94,Zuccala06invisblcoll}  would become visible and easily quantified,  including hubs in so doing.

  Two  indirect, but not to be neglected,  arguments for examining team co-authorship rather than citations stem also in the observed fact  \cite{Perssonetal04}  that in  general, co-authored publications are cited more frequently than single-authored papers. Moreover, increasingly, public and private research funding agencies require not only international and inter-institutional collaboration, but also $claim$ to search for interdisciplinary and multidisciplinary scientists, and to promote such collaborations.  To estimate the quality of such persons is far from obvious.
  
However, as mentioned, Ausloos law seems to be best for large teams, or for authors having many publications and many coauthors.  It is easily understandable, as pointed out  already in  \cite{Sofia3a}, that when an author has not many publications, or few coauthors, the law might be statistically poor. On the other hand,  deviations in presence of a large set of publications  and a large set of coauthors might be due to several reasons.  So called "intrinsic causes" might arise from the large productivity of the group based on a high turnover of young researchers, with $r>> 1$,  as well as a steady contribution from stable  partners, with $r\simeq1$. An "extrinsic reason" might arise from a large quantity of so called proceedings papers or invited lectures, on which  the list of coauthors might  be large in order to take into account various contributions on the reviewed subject and/or promote team size visibility.

 Moreover,  most prolific scientists have joint publications on different subjects. Thus, coauthors might be specific to some research subfield  of a leader. It is thus if interest to examine, for such teams and leaders, whether the law is obeyed when the research publications pertain to different subfields. Automatically, this implies to search for very prolific scientists  having worked on many different subfields.

Two cases are hereby examined. One is in fact the list of coauthors of Ausloos (MRA). He has published a little bit less than 600 papers in international journals or proceedings with reviewers. The other scientist, i.e. HES,   here below studied from the co-authorship point of view, is a guru of statistical mechanics, for which the publication list amounts to more than 1100 "papers", and for which his group website distinguishes between subfields.

After some brief introduction of the so called state of the art, in Sect. \ref{sec:artstate}, the methodology is explained in Sect. \ref{sec:Method}. The data analysis of the subfield co-authorship features is reported in Sect. \ref{sec:dataanal}, for both MRA and HES.    
In Sect. \ref{sec:dataset}, some discussion on the statistical mechanics  aspects of these illustrative cases are presented in line with general considerations on "sub-cores" of  coauthors, in Sect. \ref{sec:conclusions}. 
In  Appendix A, "small (in terms of the number of relevant  publications) subfields" are considered, combining them into a larger subfield, such that  the process can mimic the combination of subfields into the overall research field of a scientist, at  different scales.

N.B.  It will appear that  data fitting  is performed as is usually done in physics, namely a least squares fit of log-log data of the rank-frequency form. Yet, in Informetrics one prefers to fit the equivalent size-frequency form using a maximum likelihood fit. The Informetrics approach is certainly the better one\footnote{quoting an anonymous reviewer}. The approach as used here seems however  "good enough". Since each rank-frequency form has an equivalent size-frequency one \cite{Egghebook05,Adamic,AdamicHuberman}, one could indeed (have redrawn all figures from the original to the revised version of this paper. For the sake of simplicity, saving time and energy  two arbitrary chosen case  have been used for comparing the methods, and subsequent result. This is done in Appendix B. It is (fortunately) found that the results are comparable, within reasonable error bars for the numerical values.

     \begin{table}%\label{incomexpensfits}
     \begin{center} \begin{tabular}{|c|c|c|c|c|c|c|c| l |    }
  \hline
\multicolumn{2}{|c|}{MRA }&&\multicolumn{2}{|c|}{HES   }\\
 
\hline 
\hline NoP &583   &&   NoP &1160      \\
\hline NoJP &$>$ 560 &&   NoJP & $>$ 870 \\
\hline oldest P  &1971   &&   oldest P & 1965  \\
\hline latest P& 2012   &&latest P& 2012    \\ 
\hline ToNCA & $>$ 310   &&ToNCA& $>$ 480   \\ 
\hline $m_a$&20&      &$m_a$& 25 \\ 
\hline 
\end{tabular}  \end{center}
   \caption{Summary of  data  characteristics  for publications  for MRA  and HES,  up-dated till  Dec. 12, 2012 : number of publications (NoP) and of joint publications (NoJP); oldest and latest publication (P); total number of coauthors (ToNCA);  coauthor (CA) core measure ($m_a$) [1] }\label{TablestatMRAHES}
     \end{table}

\section{State of the art }\label{sec:artstate}

Disregarding disturbing effects of multi-authorship on citation impact, as shown in bibliometric studies  
\cite{Perssonetal04}  
and the effect of multiple $co-authorship$ through the $h-$index,  as yet modified in  
 \cite{hindex,hirsch10}, let  "authors" rather than "citations" be  rather  emphasized, in order to quantify research collaboration on scientific productivity \cite{MelinPersson96,LeeBozeman}. 
 
Not much seems to have been written on the measure of teams from the coauthor number point of view \cite{egg1}. Cooperation structure, group size and productivity in research groups have been studied in a modern (quantitative) way as far as apparently  as 1985 by  Kretschmer  \cite{HK94,HKretschmer85structuresize}.   
 Estimates of the returns on  quality and co-authorship outputs  have been studied for economic academia by  Sauer  in \cite{sauer88}  and Holis in \cite{Hollis01}. 
In the medical field, the "White Bull effect", i.e.  abusive co-authorship and publication parasitism, has been emphasized by Kwok \cite{Kwok05}.  Not much more to my knowledge.

 More positively, let us mention, beside the above references, work on   the critical mass and the dependency of research quality on group size by    Kenna  and  Berche  \cite{1006.0928criticalmasskenna}. Note also White's ''Toward Ego-centered Citation Analysis''  which provides a method for identifying sets of relationships between an author and others in order to define the author's multiple social networks \cite{HDWhitegarfieldfest}, though the problem of  scientific networks    \cite{newman0303516PNAS98}   is outside the present study.  Note as well a multistep process for generating bibliometric mappings of research fields and their community structure in \cite{VeldenLagoze}.
 Last, but not least, let the review by  Sonnenwald  \cite{sonnenwald07} on  scientific collaboration terminology, concepts,  classes,  stages,  positive and negative aspects,  political and socio-economic constraints, - though without quantification, be mentioned.

\section{Methodology} \label{sec:Method}

In order to quantify Ausloos   law and verify its validity limit, two cases have been selected for several reasons. First,   the list of coauthors of Ausloos (MRA) and that of HES,   are available under different conditions: on one hand,  through web sites, on the other hand through personal contacts. For example,  MRA website  $www.ulg.ac.be/supras/groupe/Staff/ausloos.html$ gives his first 360 publications, as distributed into 8 subfields.  Other papers  are also found on $http://orbi.ulg.ac.be/$. Moreover, MRA sent me his updated full publication list, according to subfields, as requested.  Book chapters and papers subsequent to scientific presentation at various    scientific meetings are included, but books and edited proceedings are not counted.

HES publication list amounts to more than 1100 "papers", and for which his group website distinguishes between subfields. Its Curriculum Vitae \& Selected Publications, taken on  $polymer.bu.edu/hes/vitahes-messina.pdf $,  lists, among other things, like  edited books, 14 book chapters and 5 encyclopedia articles,  619 articles, in the period 1966-1999 plus   more than 490 journal articles in the period 2000-up to the end of  2012.   [Listed in rank order by citation count]. It is also claimed that HES has  supervised 104 Ph.D. Theses. Interestingly for our purpose, his CV mentions 131 Research Associates and ÒVisiting ScholarsÓ. HES seems to have the largest $h$-index among physicists ($h>112$). However, in order to consider subfields, the HES  "pre-broken list", taken from $http://polymer.bu.edu/hes/topics.html$, has been used. Its content will be discussed next, some warning being necessary.

In the present approach, in order to emphasize the co-authorship features   
within different sub-fields of a (main) author, it seems  fair to me to accept {\it a priori} the sub-fields selected by this "leader". It occurs, nevertheless, that the $same$  papers are found in $different$  subfields, in the case of HES.  There are duplications.   Let it be mentioned that the case does not occur frequently. I consider that it would be very unfair to manipulate the lists  {\it a posteriori} in order to decide 
in which subfield a paper has to be put.  It seems to me that there is no "good criterion" allowing to eliminate a specific paper from one (or several) list(s).   However, it has been noticed that, in several cases,   coauthors  necessarily have, in so doing, 2 or 3 or 4 "joint publications", instead of 1 or 2, - thus overestimating this coauthor importance, - if the sum of parts is carelessly made. On the other hand,  in the present kind of study,  this "error" in estimating the number of joint publications (NoJP) seems  weakly relevant in estimating the importance of a very frequent coauthor. His/her rank will not likely be much changed, - though the NoJP is, admittedly, overestimated.
More positively, the fact that a paper appears several times is an indication of the leader multidisciplinary activity, and of his coauthors as well.  Most annoying appears to be the lack of identity between the 2012 CV list (not broken) into subfields, and the website subfield list, which seems sometimes incomplete. Also several papers seem strangely appearing in some list; the most amazing is a 2002  paper on metal-insulator transition (MIT)  
  found in the  '"Physiology and Medicine" subfield, while another on MIT does not appear anywhere. 

 In a somewhat amazing way, it was observed that a paper on where both HES and MRA are coauthors   \cite{JGRA108KI} does not appear in any subfield list.

Great care has been taken with the misprints of coauthor names: e.g.,  Gilgor, Zaleski, Kutzarova, Buldryev, Kumer,  and Giovanbattista,   are surely Gligor, Zalesky, Koutzarova, Buldyrev, Kumar,  and Giovambattista, respectively.  Great care has also been taken concerning polish, spanish, chinese and korean names.  First  (given) names and middle names, the latter sometimes missing, have been checked: e..g., T.M. Petersen and A.M. Petersen are the same person. This manual check  has allowed to distinguish name homonyms, like Ch. Laurent and Ph. Laurent.  HES also mentions a famous paper attributed to some HFS!, - in the surface physics subfields. All such and similar misprints  have been {\it a posteriori} corrected before manually counting the authors.

In conclusion, although, the final data might still be containing some "error", most of it has been manually verified and is taken as  sufficiently reliable for the present investigation. This analyzed data  is available from the author if necessary.

\section{The data and its statistical analysis} \label{sec:dataanal}. 

The  8 subfields of publications by MRA, according to the web site $http:\\www.ulg.ac.be/supras/groupe/Staff/ausloos.html$, can be defined as

1.  Condensed matter:    Disordered or Non-magnetic Materials  (1) %electronic structure of disordered alloys, solubility limits of dilute noble metals or alkali-earth metals in alloys,

2.  Condensed matter: Magnetic Materials %Transport properties of magnetic metals, alloys, compounds, in particular in the vicinity of the critical magnetic ordering temperatures, starting from Ph. D. thesis; electrical resistivity, Seebeck, or thermoelectric power, thermal conductivity,

3.  Statistical  physics:  Liquid and Amorphous States. Meteorology. %Study of the liquid and amorphous states : thermodynamic properties, and those of magnetic fluids; equation of state near the close packed limit, phase diagram and statistical mechanics of amorphous magnetic insulators and of liquid metals,

4.   Condensed matter:   Granular Materials %Infrared active modes in clusters of spheres; size effects, absorption spectrum of powders from the general solution of Maxwell's equations; nanoparticle spectra

5.  Condensed matter:   Fractures and Surfaces% : cyclotron resonance, optical properties, fractal properties, fracture, stability limits, porous media,

6. Statistical  physics:   Kinetic Growth and  Spin Models %Physical properties of spin models, generalized static and kinetic growth models,

7.   Condensed matter:    Superconductivity % : instabilities, fluctuations, various other properties, from a theoretical, phenomenological, and experimental view point, order parameter symmetry,

8.  Statistical  physics:  Econophysics, Sociophysics  %Self-Organized Criticality against extinctions and mutations in models of evolution, phase diagrams of generalized aggregation and growth models: 
\bigskip

 Due to the rather small number of joint publications and coauthors, in subfields 4 and 5, they  are  below combined into a "5\&4"  topic, for statistical purposes. The fields can be considered to  be  a "Surface Physics" one. The overlap amounts to two authors, one being the main CA in both 4 \& 5 fields, - who keeps his $r=1$ rank after the  merging, of course.  A discussion of such a case is found in the  Appendix.
 
\bigskip

The  12 subfields of publications by HES,  according to the web site  $http://polymer.bu.edu/hes/topics.html$, are

    1. Aggregation, Snowflakes, and Viscous Fingering

    2. Statistical Physics and Neuroscience (Alzheimer's Disease)

    3. Barkhausen Effect and Microfracture

    4. DNA

    5. Econophysics \& Social Science

    6. Granular Materials

    7. Physical and Social Networks

    8. Percolation, Geometric Phase Transitions

    9. Phase Transitions and Critical Phenomena

    10. Physiology and Medicine

    11. Surface Physics and Chemistry

   12.  Water
 
\bigskip

 Due to the rather small number of joint publications and coauthors, in subfield 3,  it has been  combined with subfield 6,   into a "6\&3"  topic, for statistical purposes.  The fields can be considered to  be  a "Surface Physics"  one again. The overlap amounts to two authors, whose rank is modified through the merging.

 The data is summarized in Table 2 and Table 3 for MRA and HES, respectively. Recall that the joint publications (JP) are put in different subfields, see Tables.  The number of different coauthors  (NoDCA) is given with the NoJP with the most frequent ($mf$) coauthor ($r=1$) depending on the subfield, i.e., NoJPmfCA. The number of CA having only one paper with the leader  and  the total number of coauthors   are given as  NoJP1CA 	and TNoCA respectively. 
Of course,  NoJP with only 1CA  is equivalent to the number of JP   for such authors, with  the main researcher. The total number of coauthors in a list, TNoCA, is also reported. 
 
 The characteristics of the relevant distributions are also given in Table 2 and Table 3. Statistical notations  to read the statistical Tables are  standard ones, i.e. Mean ($m$), Median, RMS, Std. Dev. ($\sigma$),  Variance (Var.), Std. Err., Skewness (Skewn.), Kurtosis
(Kurt.);  $m/\sigma$ \ is also given. 
 
      \begin{figure}
\centering
  \includegraphics[height=20.8cm,width=16.8cm]{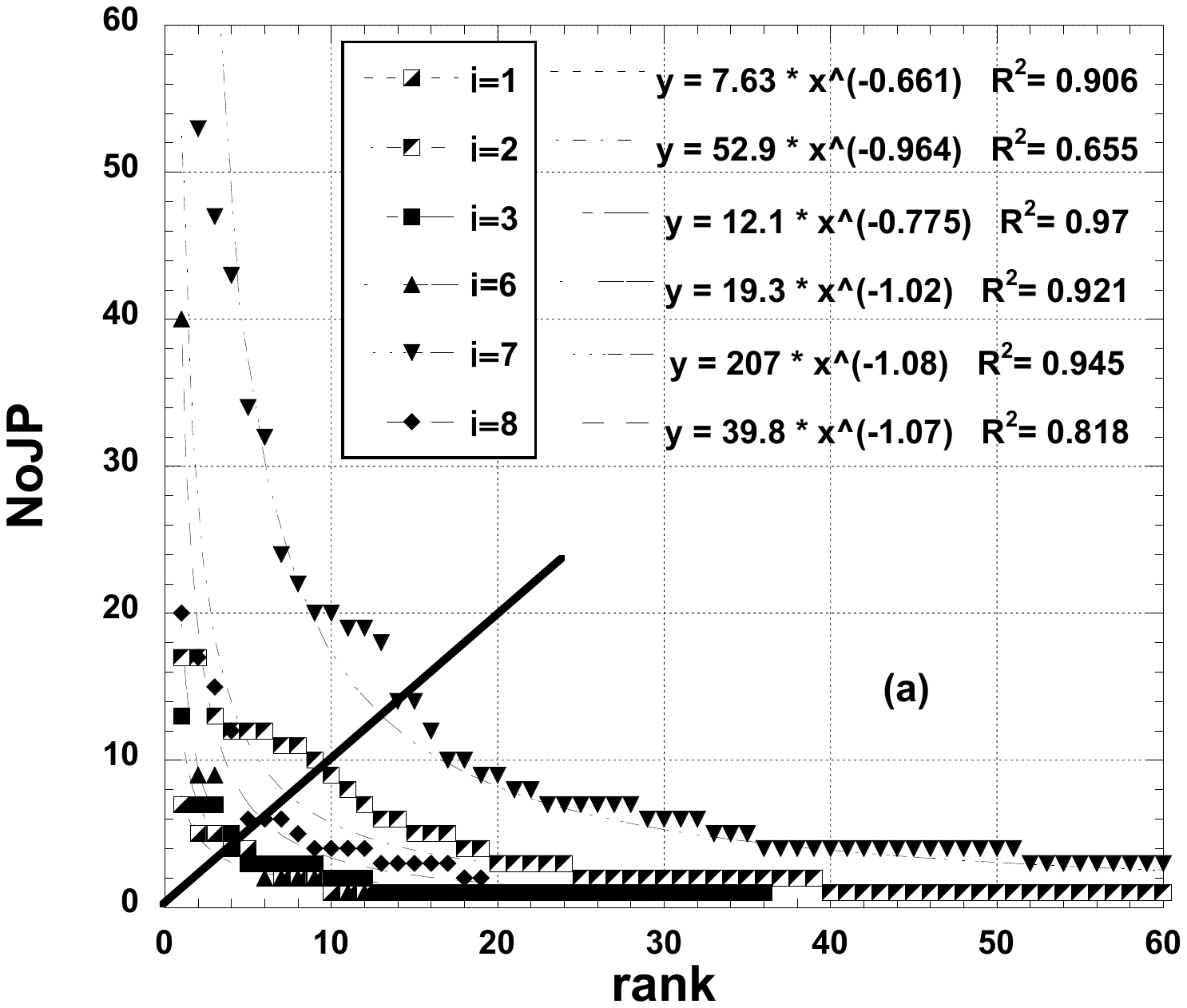}
\caption{  Number of joint publications (NoJP) for MRA, with coauthors ranked by decreasing importance for 6 different subfields ($i$; see text for $i=$ 1, ..., 8 :  (a)  in the vicinity of the  so called Ausloos coauthor core measure  \cite{Sofia3a};  (b)  log-log scale display  of  (a); the best fits are given    for the  overall range }
\label{fig:MRAsub1-8a}
\end{figure}
   
         \begin{figure}
\centering
 \includegraphics[height=20.8cm,width=16.8cm]{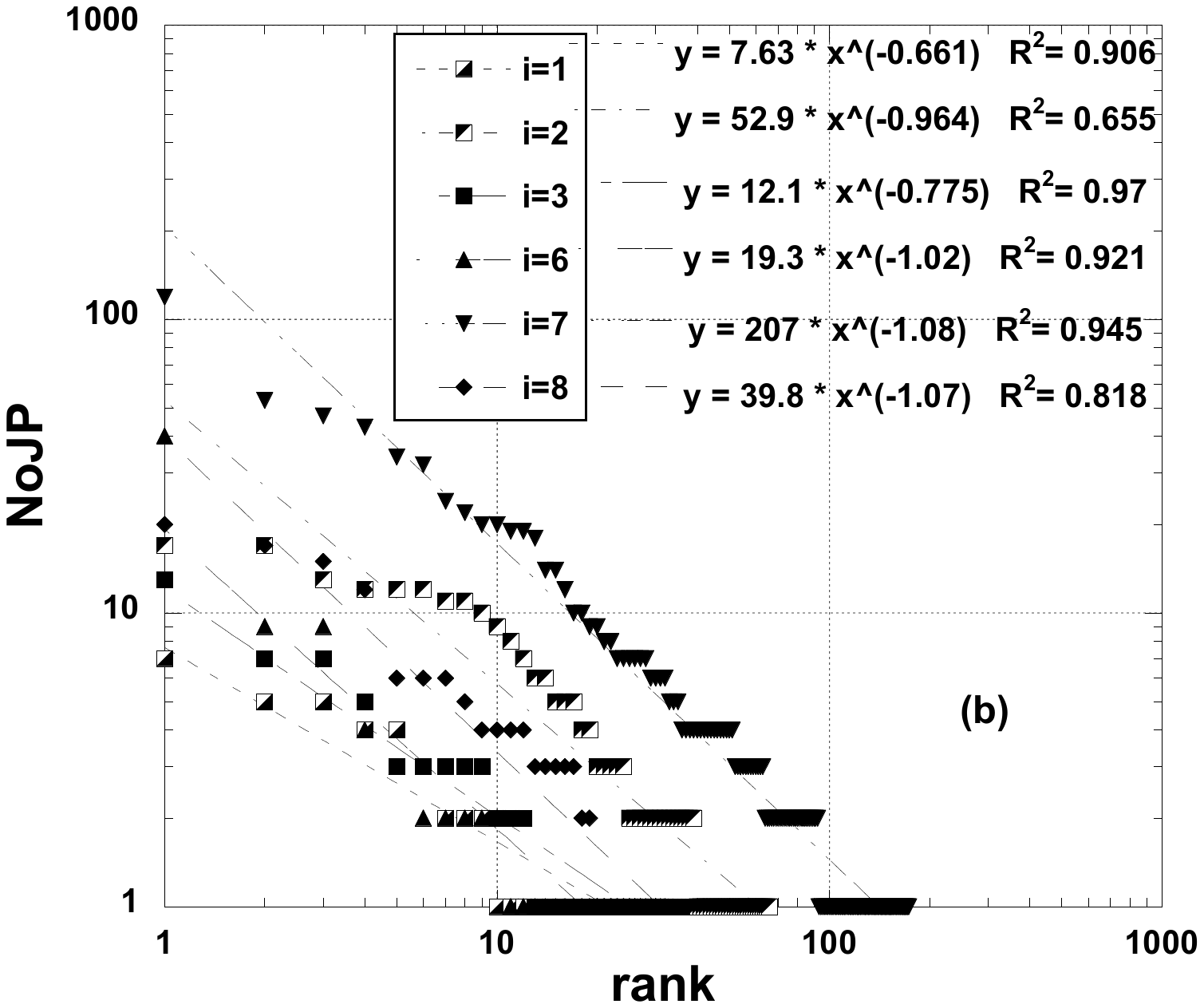}
\caption{  Number of joint publications (NoJP) for MRA, with coauthors ranked by decreasing importance for 6 different subfields ($i$; see text for $i=$ 1, ..., 8 :  (a)  in the vicinity of the  so called Ausloos coauthor core measure  \cite{Sofia3a};  (b)  log-log scale display  of  (a); the best fits are given    for the  overall range }
\label{fig:MRAsub1-8b}
\end{figure}

         \begin{figure}
\centering
  \includegraphics[height=18.8cm,width=16.8cm]{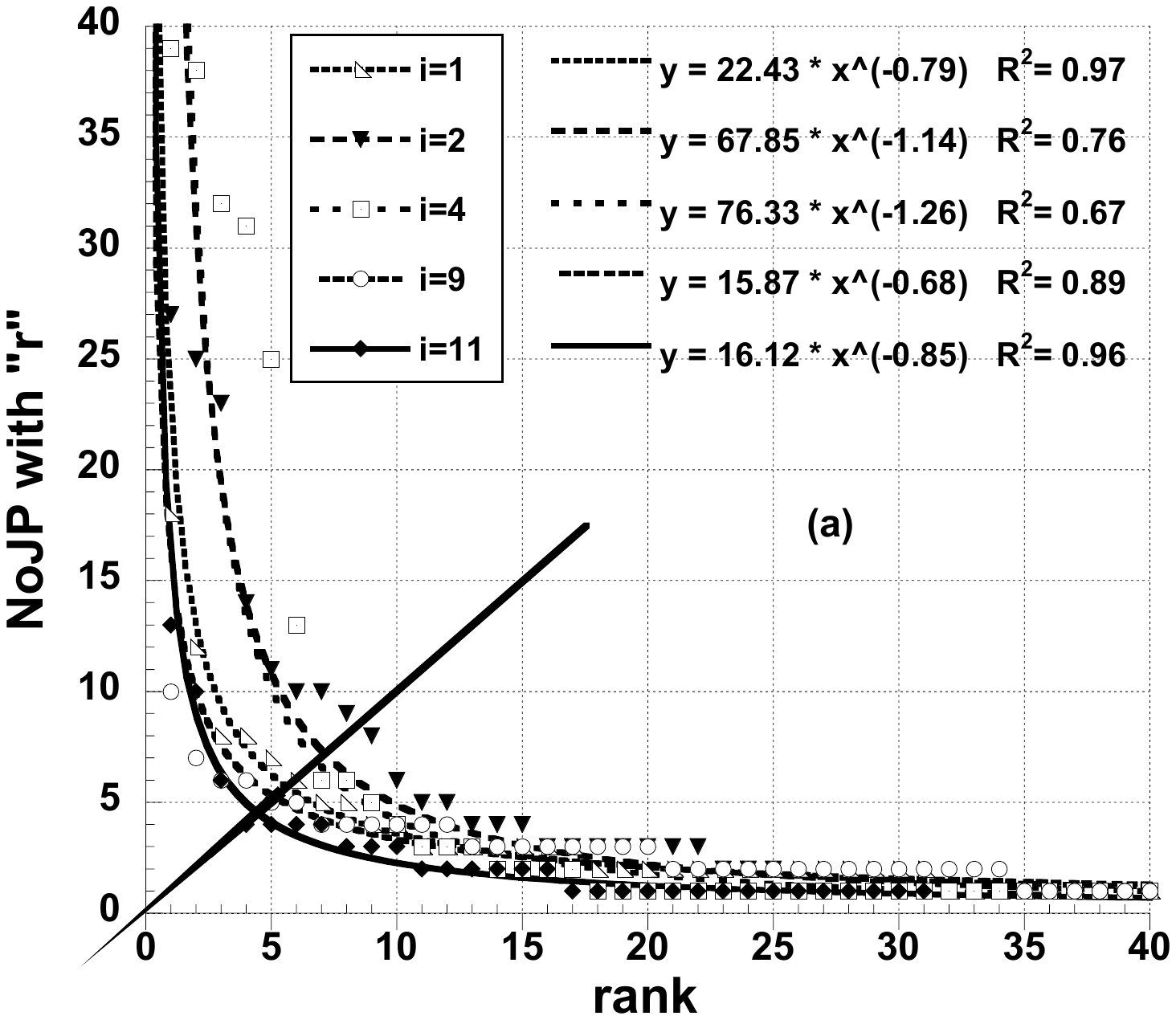}
\caption{  Number of joint publications (NoJP) for HES, with coauthors ranked by decreasing importance for the 5 "less prolific"  subfields ($i$; see text for $i=$ 1, 2, 4, 9, 11:  (a)  in the vicinity of the  so called Ausloos coauthor core measure  \cite{Sofia3a};  (b)  log-log scale display  of  (a); best fits are given    for the  overall range }
\label{fig:HESsubsmalla}
\end{figure}
         \begin{figure}
\centering
 \includegraphics[height=18.8cm,width=16.8cm]{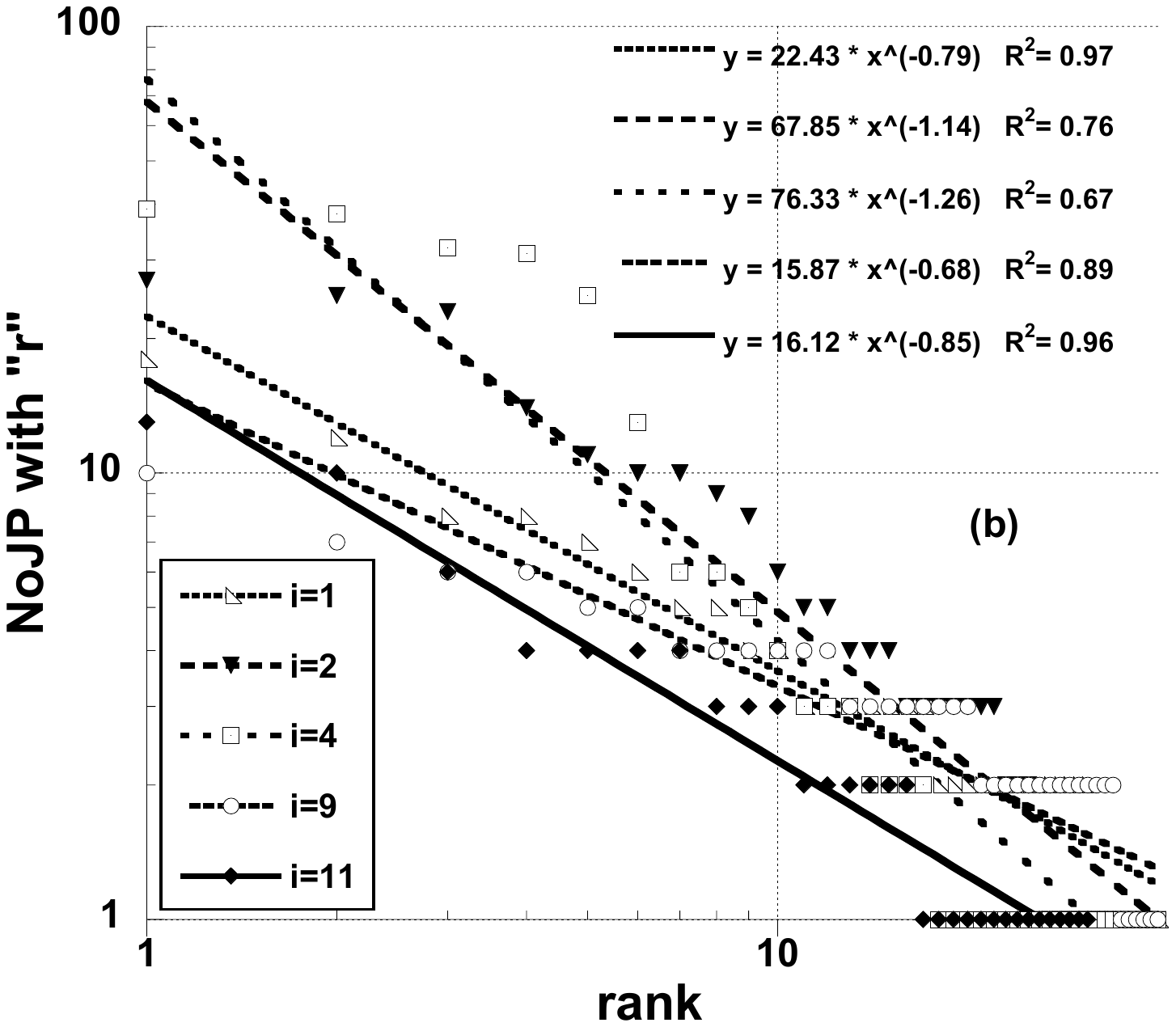}
\caption{  Number of joint publications (NoJP) for HES, with coauthors ranked by decreasing importance for the 5 "less prolific"  subfields ($i$; see text for $i=$ 1, 2, 4, 9, 11:  (a)  in the vicinity of the  so called Ausloos coauthor core measure  \cite{Sofia3a};  (b)  log-log scale display  of  (a); best fits are given    for the  overall range }
\label{fig:HESsubsmallb}
\end{figure}
    
         \begin{figure}
\centering
  \includegraphics[height=18.8cm,width=16.8cm]{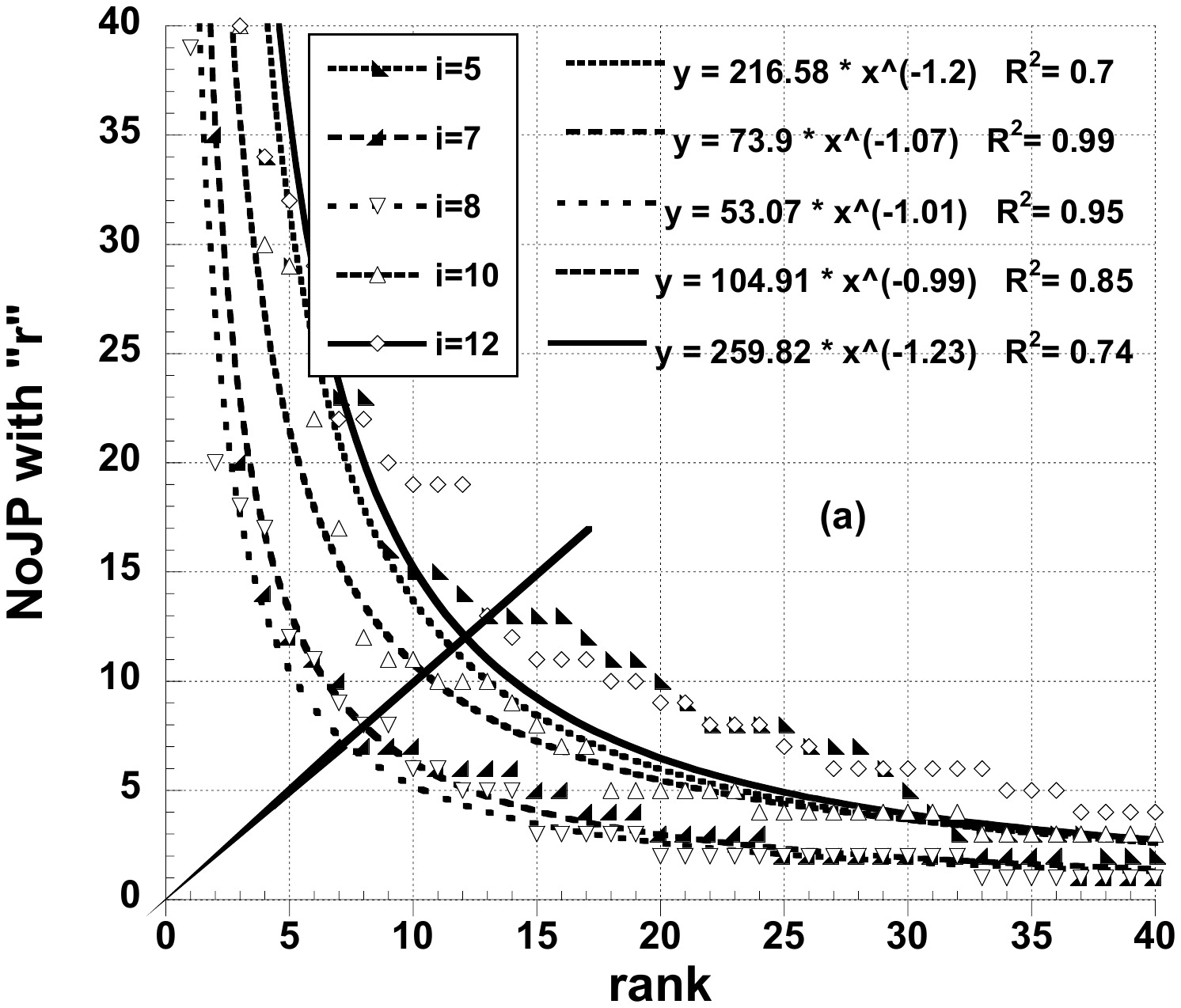}
\caption{  Number of joint publications (NoJP) for HES, with coauthors ranked by decreasing importance for the 5 "most prolific" subfields ($i$; see text for $i=$ 5, 7, 8, 10, 12:  (a)  in the vicinity of the  so called Ausloos coauthor core measure  \cite{Sofia3a};  (b)  log-log scale display  of  (a); the best fits are given    for the  overall range }
\label{fig:HESsublargea}
\end{figure}
         \begin{figure}
\centering
  \includegraphics[height=18.8cm,width=16.8cm]{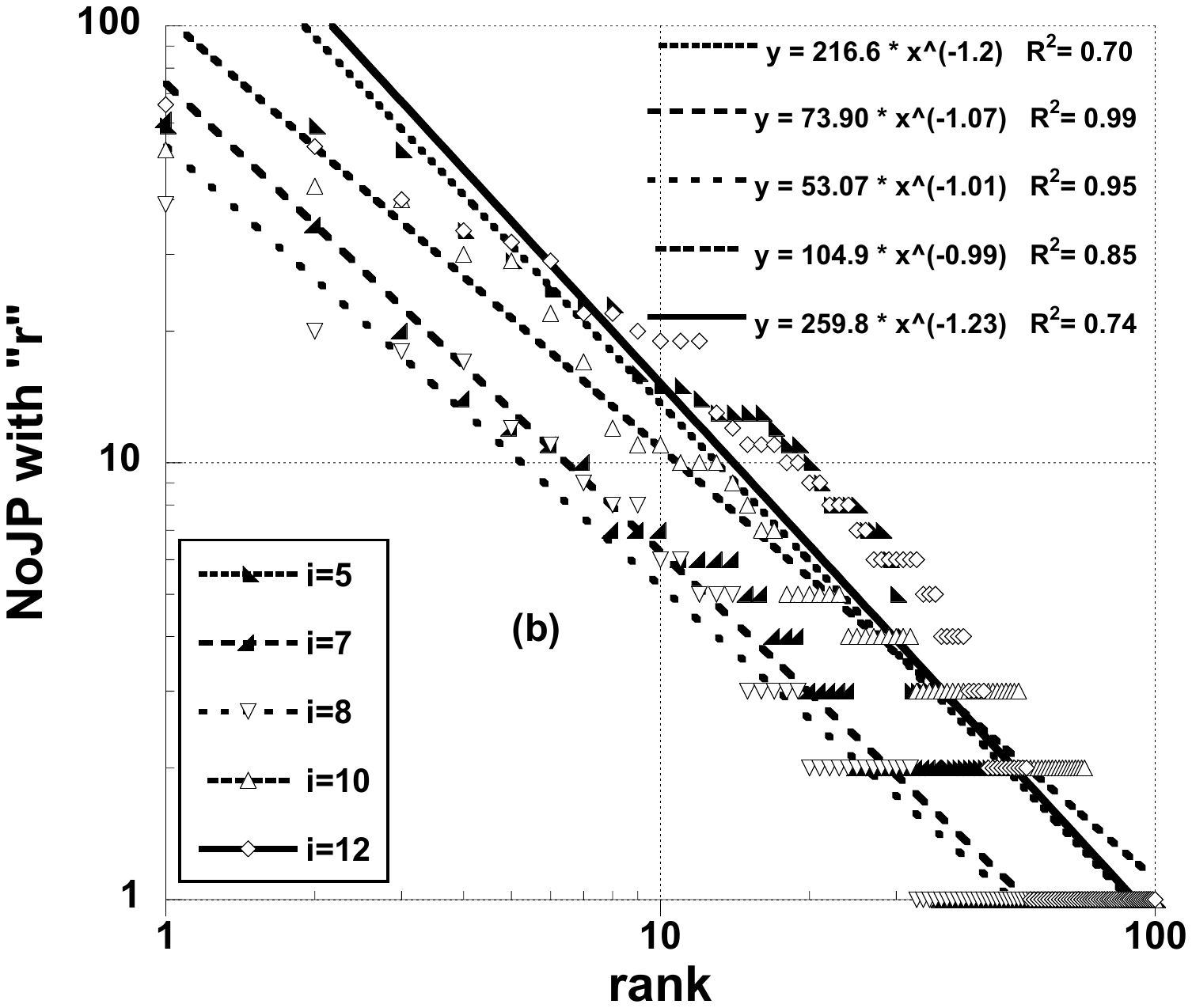}
\caption{  Number of joint publications (NoJP) for HES, with coauthors ranked by decreasing importance for the 5 "most prolific" subfields ($i$; see text for $i=$ 5, 7, 8, 10, 12:  (a)  in the vicinity of the  so called Ausloos coauthor core measure  \cite{Sofia3a};  (b)  log-log scale display  of  (a); the best fits are given    for the  overall range }
\label{fig:HESsublargeb}
\end{figure}

   	%Nottransposed 
\begin{table} \begin{center} 
\begin{tabular}[t]{cccccccccc} 
  \hline 
   $i=$   & 1  &2 & 3&4 &5 &5$\&$4&6&7& 8   \\ 
\hline   	\hline  
  oldest P	&1971	&1971&1976&1978&1978&1978&1978&1988	&1997	 	\\
latest P	&2009&2010	&2008&2006&2004&2006&2007&2011&2012	 	\\
NoJP	&24&56&30&29&7	&36	&57	&239&	95	 	\\ 
\hline 
%line220
   NoJPmfCA & 7& 17& 13&	10& 4&14& 40& 119& 20\\
NoJP1CA &23	&27&24&17&4&20 &17&81&13  \\ 
TNoCA 	&57	&246	&77	&56	&16	&72	&92	&884	&135\\  
  NoDCA &32	&66	&36&25&	9&	32&	27&	173& 32  \\
   % NoDCA &32	&66	&36&25&	9&	32&	27&	173& 32&  \hline
\hline Mean  ($m$)&	1.781&	3.727&2.139&	2.24& 1.778&	2.25& 3.407&	5.110&	4.219\\
Median &	1	&2	&1	&1	&2	&1	&1	&2	&3\\
RMS &	2.339&	5.540&	3.219&	3.225&	2.0&	3.446&	8.219&	12.771&	6.425  \\
Std. Dev. ($\sigma$) &1.539&4.131&	2.440&	2.368&	0.972&	2.652&	7.622& 11.738	&4.924\\
 $m/\sigma$ &1.157&0.902 &0.877 &     0.946 & 0.873&0.848&0.447& 0.435&	0.857  \\
Var. &	2.370&	17.063&	5.952&	5.607&	0.944&	7.032&	58.097&	137.77&	24.241 \\
Std. Err. &	0.272&	0.508&	0.407&	0.474&	0.324&	0.469&	1.467&	0.892&	0.870  \\
Skewn.  &1.990&	1.727&	2.998&	2.110&	1.320&	3.174&	4.346&	6.417&	2.003  \\
Kurt. &	3.022&	2.020&	9.563&	3.690&	1.077&	10.656&	18.247&	52.552&	3.085\\
\hline
 $m_{a}^{(i)}$&4	&9 	&4	&4 &2	&4	&4	&14&6   \\ \hline
\end{tabular} 
   \caption{Summary of  data  characteristics  for joint publications of MRA according to $i=$ 1, ..., 8 sub-fields (see text, Sect. \ref{sec:dataanal});  sub-fields 4 and 5  are here combined into "5\&4" for statistical purposes; see text for statistical notations   }\label{TablestatMRA}
\end{center} \end{table} 

     \begin{table}  
\begin{tabular}[t]{lcllcllccccccccccccccccc} 
\\ \hline 
   $i=$   & 1  &2&4 &5 &7& 8& 9 & 10&11  &12\\ 
\hline  
  oldest P	&1983&1995 &1992&1994 &2000&1976&1966&1972	&1985&1979	\\
latest P	&2004&2009&2008&2012&2007&2009&	1999	&2009&1999&2012	\\
No JP  &50&40&45&187 &77&79 &68& 116 &23&181 \\ 
\hline 
NoJPmfCA &18	&27 &39	&59 &61	&39	&10	&52	&13	&66\\
 NoJP1CA  	 &30&18 &28 &60      &32   &36         &29& 63 &15&49 \\
TNoCA 	&156	&210 &244&649&301&246&144&558&81&667\\
NoDCA &61&43&45&114&68	&68	&63	&135	&31	&104\\ \hline
Mean ($m$)	 	&2.557	&4.884 &5.422	&5.693 &4.426	&3.618	&2.286	&4.13	&2.613&	6.413\\
Median	 	&2	&3 	&1	&1 &2	&1	&2	&2&	2	&2\\
RMS	 &	3.867&	8.029 &	11.478&	11.903&	9.717&	6.954&	2.884&	8.700&	3.746&	12.441\\
Std Dev. ($\sigma$)&2.924&	6.448 0&	10.230& 10.500& 8.715& 5.983&	1.773&	7.684&	2.729&	10.712\\
 $m/\sigma$ &0.874&0.757 &0.53& 0.542 &0.508& 0.605&	1.289&0.537&	0.964&0.599 \\
Var.	 &	8.551&	41.581 &	104.66&	110.25&	75.950&	35.792&	3.143&	59.042&	7.445&	114.75\\
Std Err.	 &	0.374&	0.983 &	1.525&	0.983&1.057&	0.726&	0.223&	0.661&	0.490&	1.050 \\
Skewn.&	3.290&	2.254 &	2.431&	3.482&4.857&	3.842&	1.977&	4.208&	2.517&	3.237 \\
Kurt.	 &	12.611&	4.404 1&	4.398&	13.219&26.401&	17.404&	4.728&	18.924&	6.239&	11.991 \\
\hline
 $m_{a}^{(i)}$&6 &8 &6 &13 &7  &8	&5 &10 &	4 &13 \\\hline
\end{tabular} 
   \caption{Summary of  data  characteristics  for joint publications of HES according to $i=$ 1, ...12 subfields  (see text, Sect. \ref{sec:dataanal});  subfields 3 and 6  are not shown here, but are  combined into "6\&3" for statistical purposes and reported in Table 4 in Appendix; see text for statistical notations  }\label{TablestatHES}
%\end{center} 
     \end{table}
 
\section{Analysis and Discussion of the data set }\label{sec:dataset}

   The data is displayed and numerically fitted on Figs. \ref{fig:MRAsub1-8a}- \ref{fig:MRAsub1-8b}, 
  Figs. \ref{fig:HESsubsmalla} - \ref{fig:HESsubsmallb} and Figs. \ref{fig:HESsublargea}-\ref{fig:HESsublargeb},  in these two cases, grouped according to the size of NoJP,  NoJPmfCA, and the number of ranks for better visibility; thus in  Figs. \ref{fig:HESsubsmalla}-\ref{fig:HESsubsmallb}, one finds the $i=$ 1, 2, 4; 9, 11 subfields, and in  Fisg. \ref{fig:HESsublargea}-\ref{fig:HESsublargeb}, the $i=$ 5, 7, 8, 10, 12 subfields. 
   
      The case of  subfields $i=$ 4 and 5, for MRA, and of  subfields $i=$ 3 and 6, for HES, are treated in the Appendix.
   
   \subsection{Numerical Analysis}
   
   First, observe whether the hyperbolic law is obeyed or not. In the MRA case,  Figs. \ref{fig:MRAsub1-8a}-\ref{fig:MRAsub1-8b},  the $R^2= 0.655$ value should be considered as pretty low; it occurs  for the $i=$2 case.  The $R^2$ value is not large, but falls within usual acceptable range in this kind of studies, with non-laboratory taken data, for $i=8$. However the $R^2$ value is quite high for the other cases, i.e. $i= 1$, 3, 6, 7. In these  cases, the exponent is even quite close to +1 for the $i=$ 6 and 7 subfields. Note that the exponent is close to +1 as well for $i=$8, - the most recent subfield for investigations, see Table 2. It should be noted that the $i=2$ and 7 cases are those having the largest NoJP and NCA. 
   
   In the HES case, Figs. \ref{fig:HESsubsmalla}-\ref{fig:HESsublargeb},   the $R^2$ value should be considered as pretty low for the $i=$2 and 4 cases, and barely acceptable for the  $i=$ 5 and 12 cases.  The exponent $\alpha$  is close to +1 ($\pm 0.25$), in almost all cases, except for  $i=9$  where it is $\simeq 0.68$. %y = 146 * x^(-1.25)   R\u2\n= 0.983 
   
   In either case, it can be observed that the   $m_{a}^{(i)}$ values  are rather small and all fall much below the   overall $m_{a} $ coauthor core value. 
   
   \subsection{Influence of Subfields 
Content over Coauthor rRanking}
   
   The anomalous behavior of the $i=$2 case in MRA can likely be traced back to the (time) distribution of the publications. Indeed, there are two regimes in such a list. The first one pertains to the study of magnetic phase transitions and critical exponents through measurements and subsequent analysis of transport properties.  This leads to a large list of (portuguese) coauthors  (6 in fact, led by J.B. Sousa, with equivalent NoJP): crystal growth chemists, experimental physicists, and theoretical physicists. One obtains a so called "queen effect" \cite{Sofia3a} indicated by a sort of horizontal line in the data, see Fig. 4.  A   hyperbolic  Bradford-Zipf-Mandelbrot-like law, 
   
   \begin{equation}\label{ZMlikeCr}
  J =\frac{J^{*}}{(\nu+r)^\zeta},
  \end{equation}
  with $\zeta\simeq1$,
  might have to be considered for such cases  \cite{FAIRTHORNE}. The second regime pertains to more recent work on colossal magneto resistance. The first three coauthors being on the contrary responsible for the so called "king effect"  \cite{EPJB2.98.525stretchedexp_citysizesFR}, i.e. a sharp upturn at low $r$ values (here, $r=1,2,3$). In fact, the tail of the NoJP $vs.$  $r$ (for $r>10$) gives a remarkable hyperbola with  $R^2$ = 0.98, and $\alpha =1.25$.A huge king effect is seen for the $i$= 6 and 7 cases, with two different mfCA, i.e.,   Vandewalle and Cloots, respectively.
   
   In the case of HES, one also encounters a king and a queen effect, in several cases. The king effect is due to Gopikrishnan, Plerou and Amaral in the $i=5$ case, to Havlin in cases 1,  7,  and 8, and to Ivanov in case 10.  According to the subfield definitions, the cases are much concerned with medical topics. It is understandable that a team effect, with kings and queens, see $i=4$, 2,  and 10, are to be expected in such domains.
   
   Also observe that the case with the "worse" exponent, i.e. rather away from +1, corresponds to the oldest subfield of investigations by HES, see Table 3.
   
   Finally,    the   $m_{a}^{(i)}$ values falling much below the   overall $m_{a} $ coauthor core value can be interpreted as being due to the fact that most of the subfield core CA occur in several subfields, boosting their role in the measurement of the main author core of coauthors. These also have much sub-field disciplinarity to show on their CV.
   
  \section{Conclusions}
\label{sec:conclusions}

 An old question is : "What is measured through co-authorships?"  \cite{Laudel}  
        Indeed,  if it is possible to establish ranks between scientific products and other empirical facts,  like citations and (joint) publications,  as Beck \cite{beck84b}  discussed, it seems that only scientific achievements equal in "epistemological rank" might be admitted for statistical counts \cite{PriceModelSgrowth2004}, in order to measure some value of a scientist or a team. A test can be made if one  breaks a  ranking list into sublists and observe regularities and irregularities. The more so if the ranking is modified when breaking the list according to "inner criteria" or "intrinsic parameters".

Of course, one could warn that  the statistical methods based on mere arithmetic counts at the aggregate-level are inadequate for  at least two reasons: a quantitative bias omits relevant qualitative features and, due to its simplicity, the counting is insensitive to interactions and contextual variations  \cite{PriceModelSgrowth2004}. Moreover, duplicate papers, sometimes with only cosmetic changes,  are counted several times, and the number of coauthors seem to grow also. However, since 
 it is natural to prefer a quantitative approach, even if inexact, to any purely qualitative analysis, it is necessary to seek any data that can be obtained by a process of ''head-counting''  \cite{PriceDiscovery56}.

Ausloos coauthor core definition and measure tackles such considerations in a constructive way, through the relationship between the number ($J$) of (joint) publications with coauthors ranked according to rank ($r$) importance. A test of his findings \cite{Sofia3a}, i.e.,    $ J \propto 1/r^{\alpha}$, with $\alpha\simeq 1$,  has been made and is here above presented based on two prolific authors, i.e., having a long list of publications, and known to have many coworkers in different subfields.  Each publication list has been broken into subfields. For one, MRA,    the requested sublists have no overlap; for the second, HES,  the website lists have overlaps, but miss  a few papers, - likely outside the main subfields of interest of the scientist.
 
The effects of  data size of data  and concatenation have been studied in an Appendix, - considering  that merging of microcosmic subfields into a larger one is indicative of what can be suggested, for example,  when considering only  a large field made of several arbitrary distinguished subfields.

  Several  final  observations are to be outlined. First of all, as already pointed out by Ausloos in \cite{Sofia3a}, the simple hyperbolic law holds best for large data sets, with homogeneous distributions of NoJP and TNoCA. The effect of NoDCA entices a long tail, but  a relevant observation in the present work is the king and queen effects which force much deviation from Ausloos law at low $r$.  A Bradford-Zipf-Mandelbrot-like law, Eq.(\ref{ZMlikeCr}), might have to be tested, - with the delicate need of thereafter interpreting the two additional parameters. This is suggested for further work.

One may also conjecture that irregularities maybe due to different causes: publication inflation, proceedings counting,  co-authorship inflation, for whatever reason \cite{JASIST52.01.610lotkabreakHKRR}.   
 Moreover, it seems that  Ausloos law should be better followed for more recent investigated subfields, i.e. when NoJP is becoming large.  

Interestingly, for maybe practical considerations, a  difference in   research team behavior can be observed through the NoJP exponent and coauthor core value.   This    goes in line with the usual knowledge that scientists who collaborate  bring additional, individual goals to a collaboration as studied by Sonnenwald \cite{Sonnenwald2003a}. As she points out: a typical example is a junior scientist who wishes to be promoted and receive tenure, in addition to contributing to a collaboration. Thus individual goals  influence a scientist's ongoing commitment to a collaboration and his/her perspective on many aspects of the work \cite{sonnenwald07}. In so doing it brings much (unduly or not) influence on the co-authorship list    \cite{Kwok05}.

Nevertheless, the smallness of the core and sub-core values may imply further considerations for the evaluation  of team research purposes and activities, beside co-authorship need and necessity, within multidisciplinary aspects.
 
 The   $m_{a}^{(i)}$ values of the so called coauthor sub-core fall much below the   overall $m_{a} $ coauthor core value.   Practically, it indicates the need for globalization of measures in considering the role of the main author, and in ranking his team mates. 
   
The above analysis  also  indicates the sensitivity of the subfield notion, on one hand, and of the coauthor distribution, on the other hand, on the core measure. More work will be useful along such lines,  for better quantification. However,  this observation  can be considered to be already useful in order to imagine that one can be  introducing selection and rewarding policies in the career of members of teams,  along Ausloos coauthor core measure [1], $m_a$.

\bigskip

{\bf Acknowledgements} 

\bigskip

Thanks to M. Ausloos for private communications on    \cite{Sofia3a}, comments prior to manuscript submission and making available his  publication list broken into sub-fields.  Suggestions by   R. Cerquetti, J.M. Kowalski,  J. Mi\' skiewicz, and an anonymous reviewer - see Appendix B for this, have been welcomed.

\bigskip

     \begin{figure}
\centering
  \includegraphics[height=18.7cm,width=16.8cm]{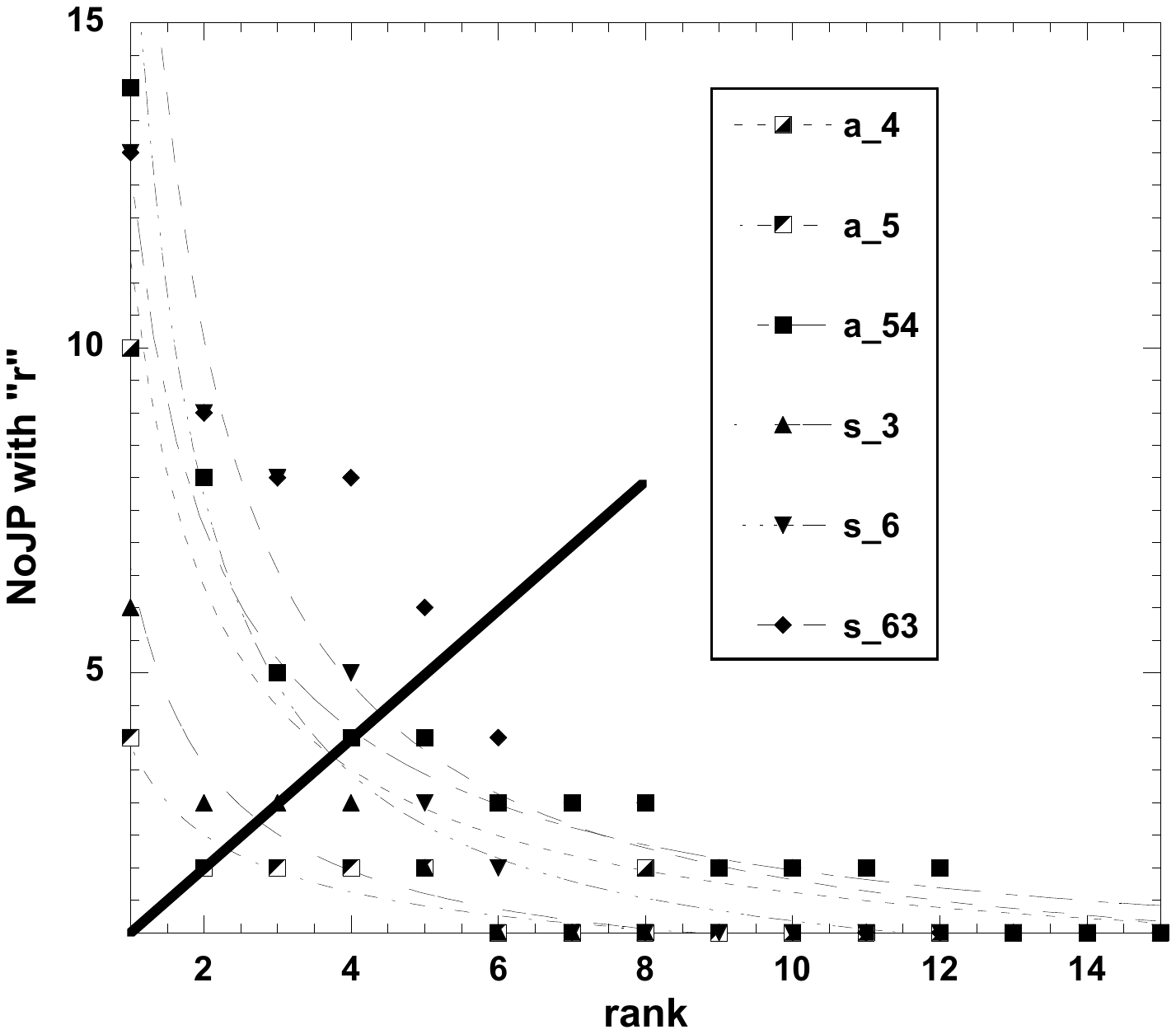}
\caption{     Number of joint publications (NoJP) with coauthors ranked by decreasing importance,  in the case of  subfields 4 and 5 (for MRA) and 3 and 6 (for HES), and their merging into a subfield "Surface physics", i.e. $5\&4$ and $6\&3$ respectively; $a_{-}4$ corresponds to $i=4$ for MRA,   $s_-3$  to $i=3$ for HES, etc.; the so called Ausloos coauthor core limit  \cite{Sofia3a} is   shown by the diagonal line; the best power law fits are given  }
\label{fig:surfa}
\end{figure}
     \begin{figure}
\centering
  \includegraphics[height=18.7cm,width=16.8cm]{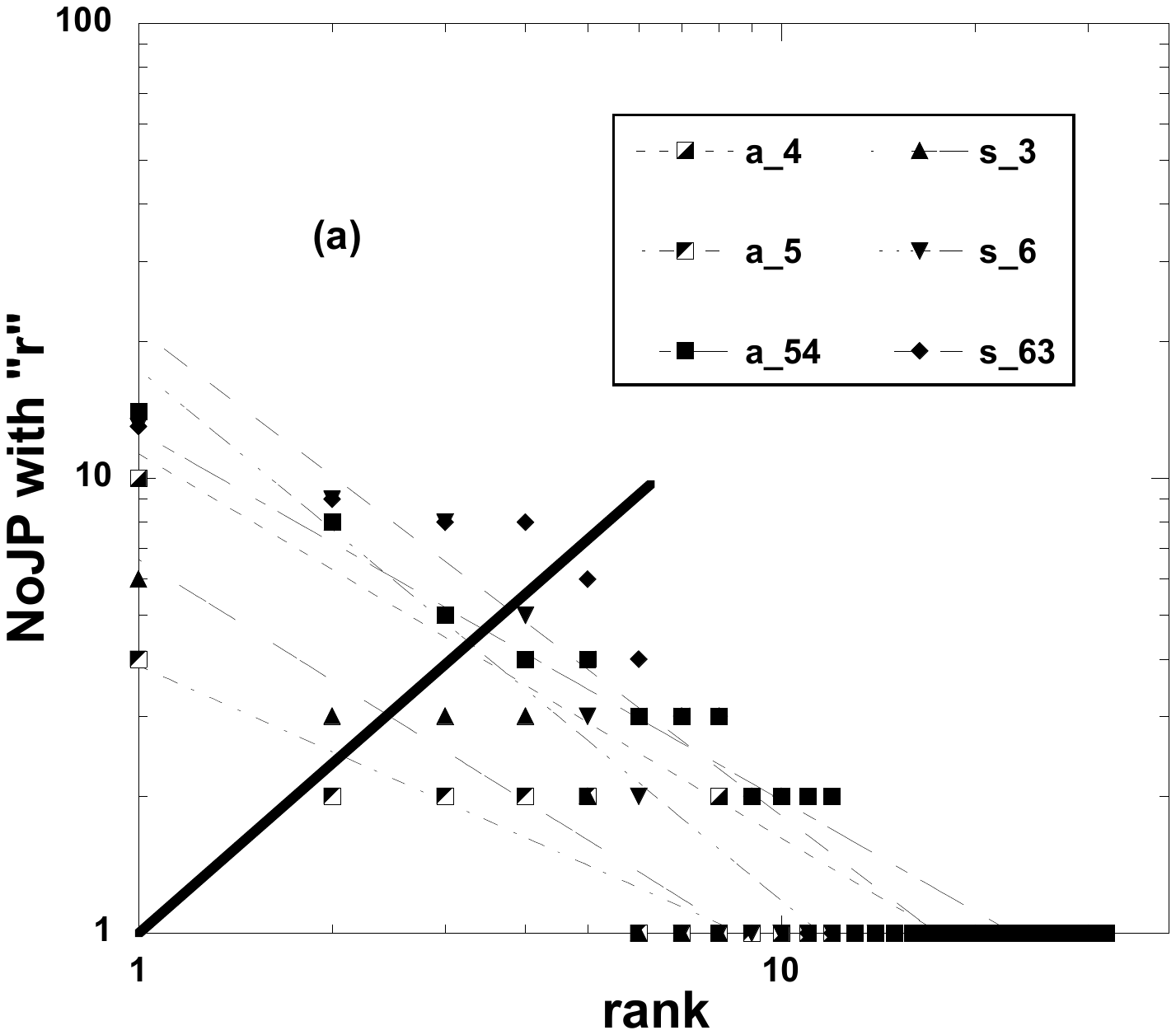}
\caption{     Number of joint publications (NoJP) with coauthors ranked by decreasing importance,  in the case of  subfields 4 and 5 (for MRA) and 3 and 6 (for HES), and their merging into a subfield "Surface physics", i.e. $5\&4$ and $6\&3$ respectively; $a_{-}4$ corresponds to $i=4$ for MRA,   $s_-3$  to $i=3$ for HES, etc.; the so called Ausloos coauthor core limit  \cite{Sofia3a} is   shown by the diagonal line; the best power law fits are given  }
\label{fig:surfb}
\end{figure}

{\bf Appendix A.   On Merging Sub-fields}

\bigskip

 \begin{table} 
     \begin{center}    \begin{tabular}{|c|c|c|c|c|c|c|c|l  |  } \hline
\multicolumn{4}{|c|}{MRA }&&\multicolumn{4}{|c|}{HES }\\
\hline
\hline   
% line 353
 $i=$  &4	&5 &  5\&4&& $i=$ &3&6	&6\&3	 \\  \hline  
  oldest P	 	&1978	&1978	&1978	&&     oldest P	 &1996 &1996& 1996	\\
latest P	 	&2006	&2004	&2006	& &latest P	 &1999 &2004	& 2004	\\
NoJP &30 & 7&  37&&   NoJP&7&15& 22 \\

\hline 
NoJPmfCA	 &	10&	4&	14&&NoJPmfCA	 &6	 	&13	&13	 \\
NoJP1CA  &17&4&20 & & NoJP1CA  	&  3  & 9 & 11    \\
TNoCA 	 	&56	&16	&72	& &TNoCA  &20		&49	&69	 \\
NoDCA	 &	25&	9&	32&	&NoDCA	 	&8	&15	  &21	 \\ \hline
Mean  ($m$)&	 2.24&	1.778&	2.25&&Mean ($m$)	 &2.5	&3.267	&3.286	 \\
Median  &1	&2 &  1&&Median	 &2.5		&1	&1	 \\ 
RMS &	3.225&	2.0 &3.446&&RMS	 &	2.958&	4.906&4.716\\
Std. Dev. ($\sigma$)  &2.368& 0.972&2.652&&Std Dev. ($\sigma$)&  1.690&	3.789&3.466\\
 $m/\sigma$ &  0.946 & 1.829&0.848& &  $m/\sigma$ & 1.479&0.862&0.948  \\	
Var.  &	5.607&	0.944&	7.032& &Var.	 & 2.857&	14.352&12.014\\
Std. Err.  &	0.474&	0.324&	0.469&& Std Err.	 & 0.598&	0.978&0.756 \\
Skewn.   &	2.110&	1.320&	3.174&&   Skewn. & 1.044&	1.516&1.4737  \\
Kurt.  &	3.690&	1.077&	10.656& &Kurt.	 & 0.31&	1.015	&1.117 \\
\hline
 $m_{a}^{(i)}$: &4	&2 &  4&& $m_{a}^{(i)}$: &3&4	&5	 \\ 
\hline
\end{tabular} 
   \caption{ Summary of  data  characteristics  for joint publications according to  merged subfields; e.g., for MRA and HES respectively, subfields 4 and 5  are  combined into "5\&4", and  subfields 3 and 6  are  combined into "6\&3" ;  see Sect. \ref{sec:dataanal} for $i=$... notations }\label{Table45MRA}
\end{center} \end{table}

In order to investigate the effect of reduced size of data in subfields, it has been mentioned that two related subfield have been merged {\it a posteriori} both in the MRA and HES cases.  The relevant data is given in Table 5;  the corresponding display is shown  in Figs.  \ref{fig:surfa}-\ref{fig:surfb}.   It is apparent that in both cases the $R^2$ value is rather large. It  increases with the data size in the MRA case, but decreases in the HES case.  In fact, the statistical data shown in Table 5 indicates   marked differences in the variance and kurtosis of the coauthor number of joint publications distribution.  In the MRA case, the NoJP1CA, TNoCA  and NoDCA are quite large.  In the HES case, the merging stretches upward theNoJP values at low rank.  This indicates a different type of research team behavior, for such "minority subfields", by the main  researchers.  I conjecture that in the former case, a more pedagogical approach is taken, frequently  involving many younger researchers; in contrast with the latter team, more prone to emphasize work by confirmed researchers.  It is observed that the sub-cores  are much smaller than the overall core. In so doing, the  $m_{a}^{(i)}$   core  value  is steady in the case of MRA but increases for HES, after merging. 
These features indicate the sensitivity of the subfield definition, on one hand, and of the coauthor distribution, on the other hand, on the core measure.

\bigskip
{\bf Appendix B.   On  rank-frequency  or  size-frequency fits}
\bigskip

An anonymous reviewer kindly pointed out that in  Informetrics one prefers to fit  data to some size-frequency  functional form using a maximum likelihood fit,  rather than  making a least squares fit  to  log-log data for the rank-frequency distribution, as more  usual in physics research.   Indeed,   according to 
Zipf's law \cite{Adamic,AdamicHuberman,Hill1,West},  in other words, the {\it rank-frequency relationship}, the  (number or) frequency  $y$  of the  occurrence of an "event" relative to  its rank $r$ follows an inverse power law, $y\sim r^{-\alpha}$. However, one can also ask \cite{Pareto}  how many times one can find 
  an  "event" greater than some size $y$, i.e. the {\it size-frequency  relationship}. Pareto's  found out that  the 
  the cumulative distribution function (CDF) of such events  follows an inverse power of $y$, or  in other words,  $P\;[Y>y] \sim y^{-\kappa}$.   Thus, the (number  or) frequency $f$ of such events   of size $y$, (also) follows an inverse power of $y$, i.e. $f\sim y^{-\lambda}$.   Some algebra \cite{Adamic}  indicates that % can convince that $(1/\lambda)+  \alpha=2$ and 
   $(1/\alpha)=\kappa$.

\begin{figure}
\centering
\caption{   Comparison of the Number of Joint Publications (NJP) as a function of the rank $r$ of coauthors (CA), and that of the    cumulative distribution function (CDF) of the number of coauthors (NCA) as a  function of NJP, for the case of  two subfields of MRA: (top)  Magnetic Materials;  (bottom) Superconductivity.  A classical  plot as well as a log-log scale display are presented with the best fits given    according to  the least-square  fit method.  The resulting numerical values and the corresponding $R^2$ values are given } 
  \includegraphics[height=19.5cm,width=16.8cm]{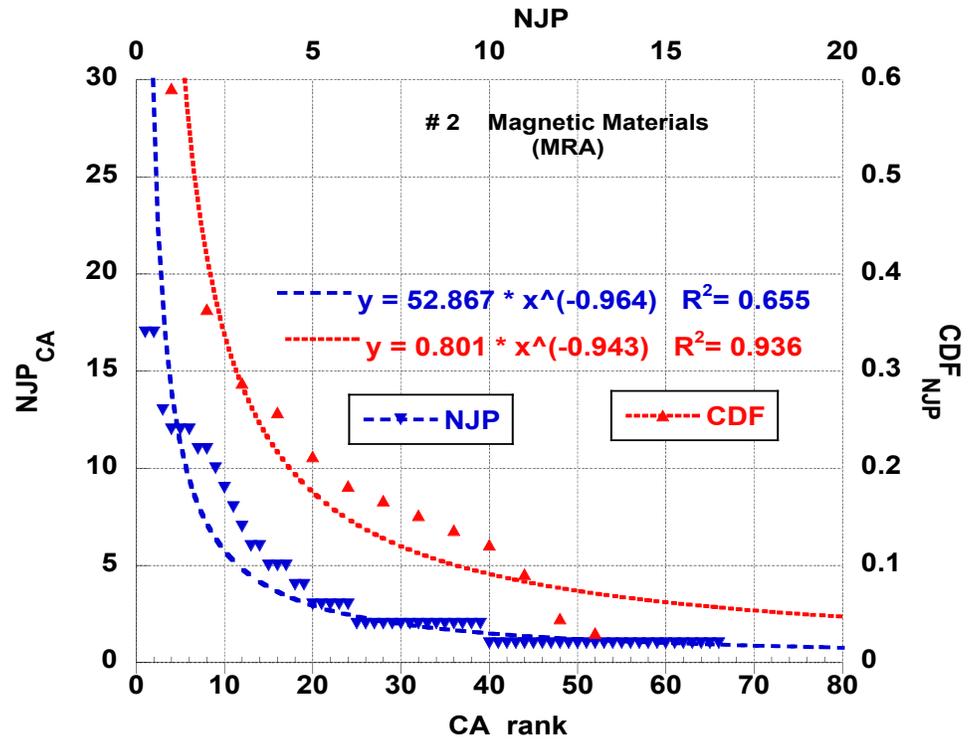}
\label{fig:zipforCDFa}
\end{figure}

\begin{figure}
\centering
\caption{   Comparison of the Number of Joint Publications (NJP) as a function of the rank $r$ of coauthors (CA), and that of the    cumulative distribution function (CDF) of the number of coauthors (NCA) as a  function of NJP, for the case of  two subfields of MRA: (top)  Magnetic Materials;  (bottom) Superconductivity.  A classical  plot as well as a log-log scale display are presented with the best fits given    according to  the least-square  fit method.  The resulting numerical values and the corresponding $R^2$ values are given } 
  \includegraphics[height=19.5cm,width=16.8cm]{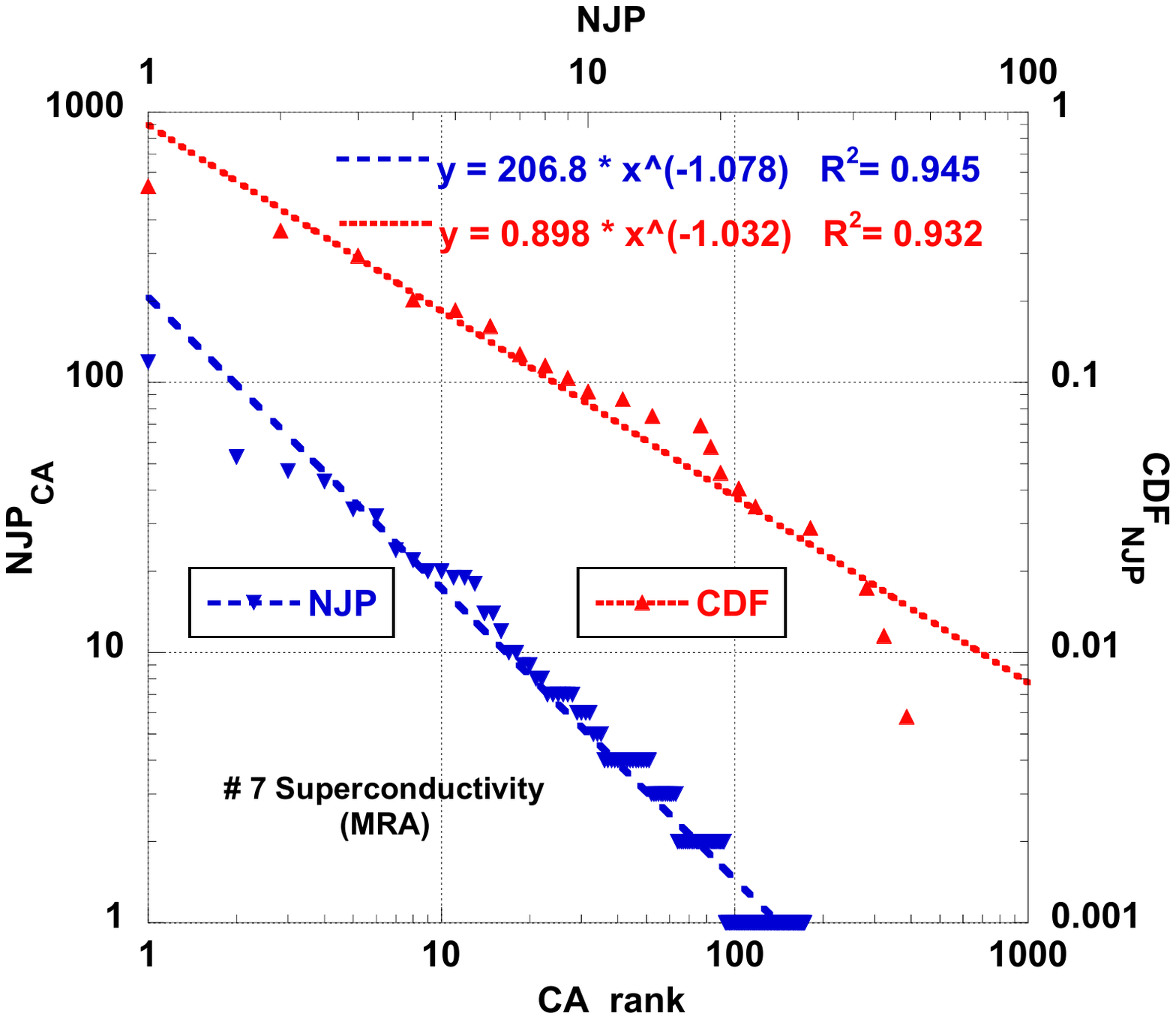}
\label{fig:zipforCDFb}
\end{figure}

Both sides of the alternative make sense. Thus, one could  redraw all figures from the main text and turn them into size-frequency plots. For the sake of simplicity, saving time and energy,  two  cases  have been quite arbitrarily chosen for comparing the methods and subsequent results. This is  presented in Figs. \ref{fig:zipforCDFa}- \ref{fig:zipforCDFb}.  The data on two subfields of MRA: (a)  Magnetic Materials  (field $\#$2) and  (b) Superconductivity (field $\#$7)  are analyzed in two ways.  
First, as in the main text,  the Number of Joint Publications (NJP) is shown as a function of the rank $r$ of coauthors (CA), in order to find the Zipf exponent. Next,   the cumulative distribution function (CDF) of the number of coauthors (NCA) is presented.   
A classical  plot is shown  as well as  a  log-log scale display in order to illustrate the methods.  The best fits are given  
  according to  the least-square fit method.  The resulting numerical values and the corresponding $R^2$ values are given. In the Magnetic Materials case, the $R^2$ value is rather low ($\simeq 0.655$), but  this  can be attributed  to the very  small number of data points, as discussed in the main text. For such a case, the CDF vs. NCA plot is much better ($R^2\simeq 0.936)$; it leads to $\kappa_2\simeq0.943$, in obvious notations.  A direct fit gives of $f(y)$ gives  $\lambda_2\simeq1.085$ ($R^2 
 \simeq$   0.947).
%  Note that    $(1/\lambda_2)+\alpha_2 \equiv 0.922+0.964=1.886 \sim 2$, - in obvious notations.  
   The exponent $\alpha_2\simeq 0.964$ seems thus reliable, accepting the error bars. 
    Concerning the log-log plot display (Fig.6) for the case of the papers on Superconductivity,  $R^2$ values are fine
($\sim$ 0.945 and 0.932) for $\alpha_7\simeq 1.078$ and
% thus $1/\lambda\simeq1.042$; and
$\kappa_7\simeq1.033$,   respectively, while $\lambda_7\simeq0.96$ ($R^2 
  \simeq$   0.932).
% the exponent sum leads to  $(1/\lambda_7)+\alpha_7 \equiv 1.042+1.078=2.120 \sim 2$.  
 
 If one wishes to use the maximum likelihood method \cite{SIAM51.09.661powerlaws,Egghebook05,Mitchener}, but assuming for  comparison  with the above that the optimal function is a mere power law, one finds  using the Table 1 of  \cite{RousseauJD49.93.409} (reproduced in  \cite{Egghebook05}) in order to estimate the exponent of interest from the ratio $-\zeta'/\zeta$: 
  $\alpha_2\simeq 1.357(0.051)$, $\lambda_2 \simeq 1.76(0.912)$, 
 and $\kappa_2\simeq 1.745(0.132)$; on the other hand, 
 $\alpha_7\simeq1.338(0.823)$,  $\lambda_7 \simeq 1.775(0.98)$,  
and $\kappa_7\simeq 1.665(0.474)$, where in the  (...) is given the corresponding $R^2$.
 %leading to  
% $\lambda_2+\alpha_2 \equiv 0.566+1.359=1.925 \sim 2$ and  
%to $\lambda_7+\alpha_7 \equiv 0.560+1.338=1.898 \sim 2$. 

As expected, the results are thus  different, but often comparable, within reasonable error bars.  Of course, other theoretical laws can be examined. The exponential and the logarithmic forms have been tested for these cases. They lead to so  bad results (sometimes $R^2<0.2$) that they are not shown;  no other  functional form has been investigated. The case of a generalized Pareto distribution \cite{FAIRTHORNE,MandelPareto}, as mentioned in the main text, is left for further work.    In any case, however, the possible power law  relations  seem to indicate that  co-authors do  form  clusters which  are  locally 
scale-free, within the overall scientific  network     \cite{newman0303516PNAS98}.

\end{document}